# Orthogonal Flatbands-Enabled Robust Fano Resonances through Brillouin-Zone Folding

*Yangsong Ye*[1,2], *Hanchuan Chen*[2], *Shijie Kang*[2], *Haitao Li*[2], *Ken Qin*[2], *Boyuan Ge*[2], *Xiexuan Zhang*[2], *Jingrong He*[2], *Jiusi Yu*[2], *Xiaoxiao Wu*[2,*], *Chunlei Yang*[1,*]

[1]Center for Photonics Information and Energy Materials, Shenzhen Institute of Advanced Technology, Chinese Academy of Sciences, Shenzhen 518055, China

[2]Modern Matter Laboratory and Advanced Materials Thrust, The Hong Kong University of Science and Technology (Guangzhou), Nansha, Guangzhou 511400, China

a) Corresponding authors. Electronic mail: xiaoxiaowu@hkust-gz.edu.cn (X. Wu), cl.yang@siat.ac.cn (C. Yang).

**Abstract:** Fano resonances based on coherent interference between localized Lorentz modes and Fabry–Pérot radiation continua have been widely explored for microwave metamaterial sensors, filters, and photonic devices. However, conventional Fano metasurfaces suffer from angular dispersion and polarization-dependent spectral variations, limiting robustness under dynamic illumination. Here, we propose a band-folding-enabled flatband engineering strategy to realize robust Fano resonances in a planar microwave metasurface with a 2×2 enlarged cell of orthogonally arranged H-shaped metallic resonators. The enlarged cell induces Brillouin-zone folding, generating a nearly dispersionless flatband near 15.3 GHz with suppressed momentum dependence and enhanced photonic confinement. This flatband state provides a high-Q localized resonance strongly coupled to the FP radiation continuum, producing a stable asymmetric Fano response. Furthermore, the orthogonal resonator configuration supports two polarization-decoupled dipole modes with comparable excitation efficiency, reducing polarization-induced spectral distortion. Near-field measurements and simulations verify localized flatband features, while far-field experiments demonstrate stable Fano responses under oblique incidence from 0° to 30° with minor resonance variation. A temporal coupled-mode theory model is developed to quantitatively describe Fano spectral evolution under different illumination conditions. This work establishes a flatband-mediated route toward robust microwave Fano metasurfaces and offers a general strategy for stable resonant photonic device design.

## 1.Introduction

Fano resonance[1–3] originates from the coherent interference between discrete eigenstates and continuous radiative continua, a phenomenon initially observed in quantum atomic scattering systems[4–6]. Unlike conventional Lorentzian resonances [7,8], Fano resonances exhibit characteristic asymmetric spectral line shapes with much narrower linewidths and stronger spectral selectivity, making them highly attractive for realizing high-quality resonant responses. Owing to these unique properties, the Fano interference mechanism has been desirably transferred from quantum systems into photonic platforms through artificially engineered photonic structures [9]. Photonic Fano resonances have subsequently become an important approach for manipulating electromagnetic (EM) waves and have been widely applied in various fields, including filters [14–16], biochemical sensors [17–19], nonlinear nanophotonics [11], subwavelength meta-optics [12], and photonic signal processing [20–22] through metamaterial and metasurface engineering [13]. In these scenarios, such as using hybrid metal-dielectric metasurfaces, Fano resonances are typically realized through coherent coupling between Lorentzian localized modes supported by metallic resonators and FP continuum modes induced by substrate-mediated multiple scatterings [7,25,26], providing an effective approach for developing high-performance photonic devices.

Despite the achievements, conventional Fano metamaterials and metasurfaces generally suffer from fabrication imperfections and variations in excitation conditions, which can induce substantial spectral distortions and instabilities. Unfortunately, these unavoidable issues are also amplified by the spectral sensitivity of Fano resonances. Topologically protected Fano structures [23,24] have been proposed to enhance modal robustness; however, their complex structural requirements and limited compatibility with planar fabrication restrict practical implementation. For commonly used FP–Lorentz coupled metasurfaces, variations in the incident angle and polarization can significantly modify the in-plane wavevector and excitation efficiency, resulting in resonance shifts and amplitude fluctuations [27]. These limitations originate from the intrinsic dispersion of conventional photonic modes and the anisotropic response of asymmetric resonators. Photonic flatbands, characterized by negligible frequency dispersion, extremely high photonic density of states (DOS) [31–33], and near-zero group velocity, provide a promising strategy for suppressing wavevector-dependent resonance variations [28–30], while supporting strong Fano resonances. Enlarged cell-based band-folding engineering can reconstruct photonic band structures and generate target-frequency flatbands through Brillouin-zone compression while maintaining compatibility with planar metasurface fabrication [34–37]. However, flatband physics and Fano interference have so far been mainly investigated independently. Existing studies have focused either on fundamental flatband phenomena, including topological transport [38,39] and slow-light effects [40], or on optimizing Fano resonance under fixed illumination conditions. Few studies have integrated band-folding-induced flatband regulation with symmetric resonator configurations to simultaneously improve the angular and polarization stability of FP–Lorentz-type Fano resonances. Therefore, developing a unified design strategy that integrates flatband engineering with Fano resonance stabilization, together with systematic investigations of its underlying physical mechanism, remains an open question.

Here, we propose a planar metasurface to realize robust Fano resonances based on orthogonal photonic flatbands through band-folding engineering. The enlarged cell, composed of orthogonally arranged H-shaped metallic resonators, induces folding of the first Brillouin zone (FBZ) and generates a pair of nearly dispersionless flatbands, which effectively suppresses wavevector-dependent resonance variation under oblique incidence. Meanwhile, the orthogonal configuration supports two polarization-decoupled dipole modes, enabling stable excitation under different polarization states and reducing polarization-induced spectral distortion. The flatband localization properties and robust Fano response are systematically investigated through eigenmode simulations, near-field microwave microscopic measurements, and far-field transmission experiments. The fabricated metasurface experimentally demonstrates stable Fano transmission characteristics under oblique incidence ranging from 0° to 30°, with only minor resonance frequency deviation and amplitude variation. Furthermore, a temporal coupled-mode theory (TCMT) model is established to analyze the origin of the Fano spectral features due to the interference between the flatbands and the Fabry-Pérot background. This work provides a general and practical strategy for enhancing the angular and polarization robustness of Fano resonances in metasurfaces through flatband engineering and offers new insights into flatband–FP interference mechanisms in hybrid metal–dielectric systems for stable filtering, sensing, and switching applications.

## 2. Results and Discussion

### 2.1. Flatband formation through Brillouin-zone folding engineering

The proposed metasurface is composed of periodically arranged H-shaped metallic patterns, with a rotational degree of freedom $\beta$ in the 2×2 enlarged cell, as illustrated in Figure 1(a). The special case $\beta = 90°$ that we will focus on is displayed separately. The geometrical parameters are optimized through full-wave simulations, with the copper linewidth, horizontal arm length, vertical arm height, lattice period, substrate thickness, and copper thickness set as $w = 0.125$ mm, $b = 1.75$ mm, $a = 3$ mm, $L = 5$ mm, $h = 3$ mm, and $t = 0.035$ mm, respectively. The corresponding numerical simulation settings, including boundary conditions and calculation parameters, are summarized in Supplementary Information (SI), Note 1. These parameters are also compatible with standard printed-circuit-board (PCB) fabrication processes, ensuring reliable experimental realization. The enlarged cell with 2×2 metallic patterns introduces a band-folding effect, which folded the original first Brillouin zone (FBZ) to only one quarter, as shown in Figure 1(b). Consequently, photonic bands of the primitive lattice are folded back and hybridized within the quarter, leading to substantial reconstruction of the photonic dispersions when $\beta$ is not equal to 0 degree. Unlike conventional metasurfaces relying solely on isolated meta-atom resonances, this band-folding mechanism provides a universal route for manipulating collective photonic states and enables the formation of nearly dispersionless flatbands. Figure 1(c) schematically illustrates the designed metasurface and its robust Fano resonance response under different incident angles and polarization states, highlighting its capability for angle- and polarization-insensitive resonant transmission.

To quantitatively clarify the flatband formation mechanism, eigenfrequency calculations are performed along the high-symmetry path of the folded FBZ of the enlarged cell. By continuously tuning the rotation angle $\beta$ (presented in SI, Note 3), the evolution of photonic bands is systematically tracked. Representative band structures and density-of-states (DOS) spectra for the $\beta = 0°$ and $\beta = 90°$ configurations are shown in Figure 1(d,e), whereas the complete evolution of the DOS with increasing $\beta$ is presented in SI, Note 4. The pronounced DOS enhancement at the flatband frequency indicates a strong accumulation of electromagnetic states and an increased photon lifetime within the metasurface. Such a high density of localized states provides the narrow resonant channel required for generating sharp asymmetric Fano interference. The rotational modulation alters the near-field coupling between adjacent H resonators, thereby continuously modifying the hybridization strength between localized resonant states and collective enlarged cell modes. This tunable mode interaction provides an effective degree of freedom for reconstructing the photonic dispersions and driving the transition toward a nearly dispersionless flatband state. The calculated band evolution reveals that $\beta = 90°$ produces a nearly dispersionless band centered at approximately 15.3 GHz, which leads to the observed Fano resonances. The electromagnetic characteristics of the engineered enlarged cell are further examined through eigenmode analysis at the Γ point. Figure 1(f) and 1(g) show the out-of-plane electric-field ($E_z$) distributions of the two orthogonal electric dipole eigenmodes ($P_x$ and $P_y$) supported by the $\beta = 90°$ enlarged cell at flatband frequency. The complementary field distributions indicate that the two dipole modes are selectively excited in orthogonal H-shaped resonators, providing the modal basis for the polarization-insensitive response discussed in the following sections.

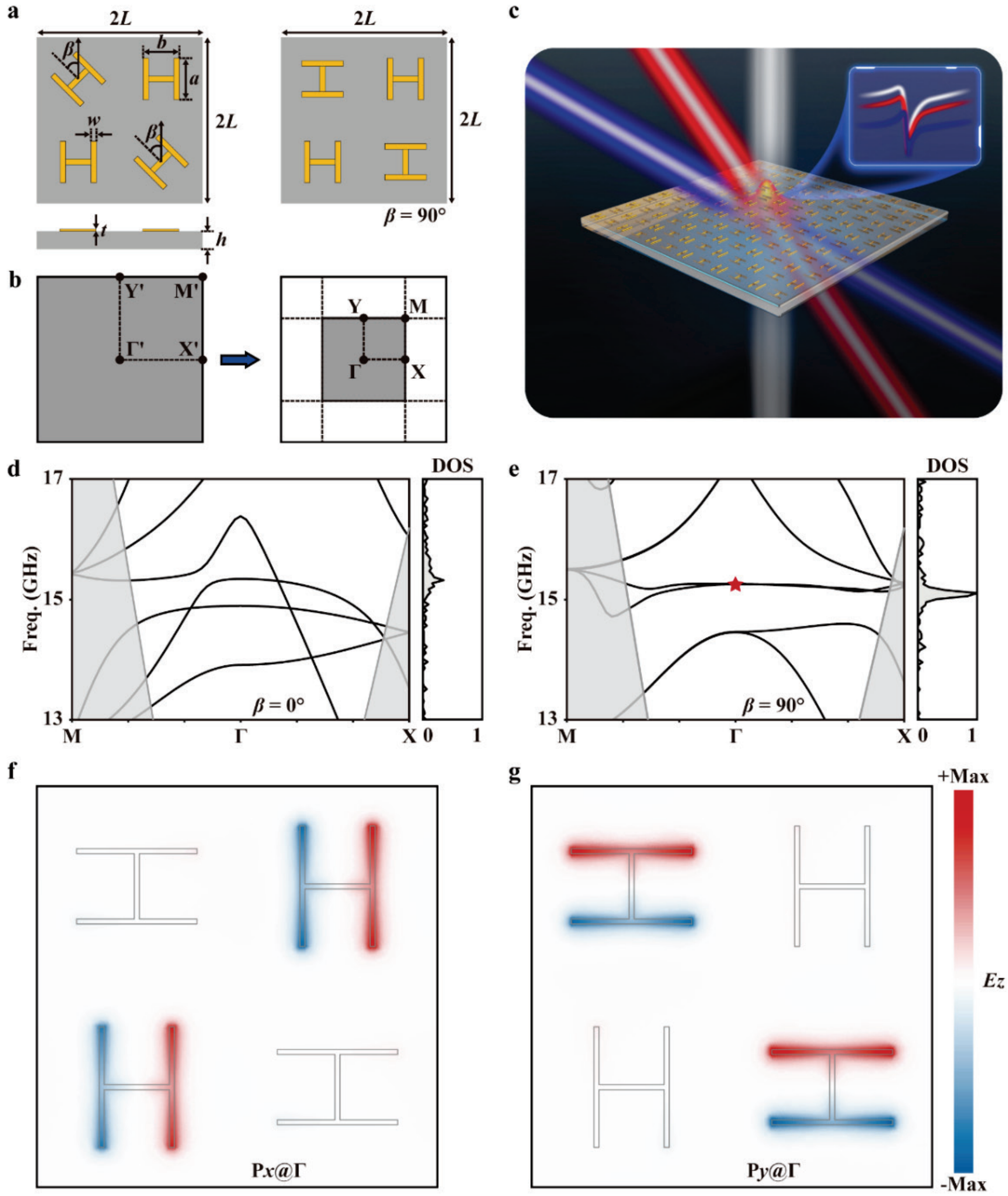

Figure 1 | Flatband engineering and robust Fano resonance in a 2×2 H-resonator enlarged cell metasurface. (a) Schematic illustrations of the designed H-resonator

enlarged cell with tunable rotation angle $\beta$ (left) and the case $\beta$ = 90°, that is, the orthogonally arranged configuration (right). The optimized geometric parameters, including lattice period (2$L$), resonator dimensions ($a$, $b$, $w$), substrate thickness ($h$), and metal thickness (t), are labeled. (b) Schematic of Brillouin-zone folding induced by lattice enlargement from the primitive 1×1 unit cell to the 2×2 enlarged cell, in which high symmetric points are marked. (c) Three-dimensional schematic of the metasurface under oblique incidences; the inset shows the asymmetric Fano transmission spectrum. (d,e) Calculated photonic band structures  and for the initial configuration $\beta$ = 0° (d) and orthogonal configuration $\beta$ = 90° (e) along the folded Brillouin-zone path M–Γ–X and corresponding DOS spectra. Red star markers indicate the representative eigenmodes associated with the flatband states when $\beta$ = 90°. (f,g) Simulated out-of-plane electric-field ($E_z$) distributions of the two orthogonal flatband eigenmodes at the Γ point for the $\beta$ = 90° configuration.

### 2.2. Experimental Verification of Flatband-Induced Electromagnetic Localization

Following the band-structure analysis in Figure 1, we experimentally characterize the near-field electromagnetic distributions to directly verify the flatband-induced localization behavior in the proposed metasurface. A scanning near-field microwave microscope is employed to map the spatial electric-field distributions of the fabricated 2×2 H-resonator enlarged cell and the detailed experimental setup is illustrated in SI, Note 2. The probe antenna is positioned 2 mm above the sample surface using a precision translation stage inside a shielded environment, while the fabricated sample maintains identical structural and dielectric parameters to those used in simulations. After background calibration using a bare substrate, reliable near-field measurements are obtained with sufficient signal-to-noise ratio. The measured field distributions at three representative frequencies (14.3, 15.3, and 16.3 GHz) are presented in Figure 2(a–c), together with corresponding full-wave simulation results in Figure 2(d–f).

The measured near-field evolution directly reveals the frequency-dependent transition of electromagnetic transport predicted by the calculated photonic bands. At 14.3 GHz, below the flatband frequency, the system remains in a dispersive regime, allowing surface waves to propagate efficiently across the metasurface plane. As shown in Figure 2(a,d), both experimental and simulated fields exhibit extended circular interference fringes surrounding the excitation source, indicating weak electromagnetic confinement and dominant wave propagation behavior. At the designed flatband frequency of 15.3 GHz, a pronounced transition from wave propagation to field localization is observed. As shown in Figure 2(b,e), the electromagnetic energy is concentrated within the excitation region, while lateral propagation is strongly suppressed. The highly localized hotspot and reduced spatial spreading directly demonstrate the enhanced confinement of the flatband mode induced by Brillouin-zone folding. At 16.3 GHz, above the flatband frequency, the system returns to a dispersive regime and exhibits directional energy transport. The cross-shaped near-field distributions in Figure 2(c,f) indicate anisotropic propagation caused by the recovered band dispersion, further confirming the strong dependence of photonic transport on the underlying band structure.

The continuous evolution from extended radial propagation below the flatband, to strong electromagnetic localization at the flatband frequency, and finally to directional wave transport above the flatband is in excellent agreement with the calculated band evolution. These observations provide direct experimental evidence that the localized electromagnetic state originates from the band-folding-induced flatband rather than from isolated resonances of individual H-shaped resonators. The experimentally verified flatband localization establishes the physical foundation for maintaining a stable high-Q resonant state, which is essential for robust Fano interference. Based on this confirmed flatband platform, the following sections further investigate its influence on the angular and polarization stability of the far-field Fano transmission response.

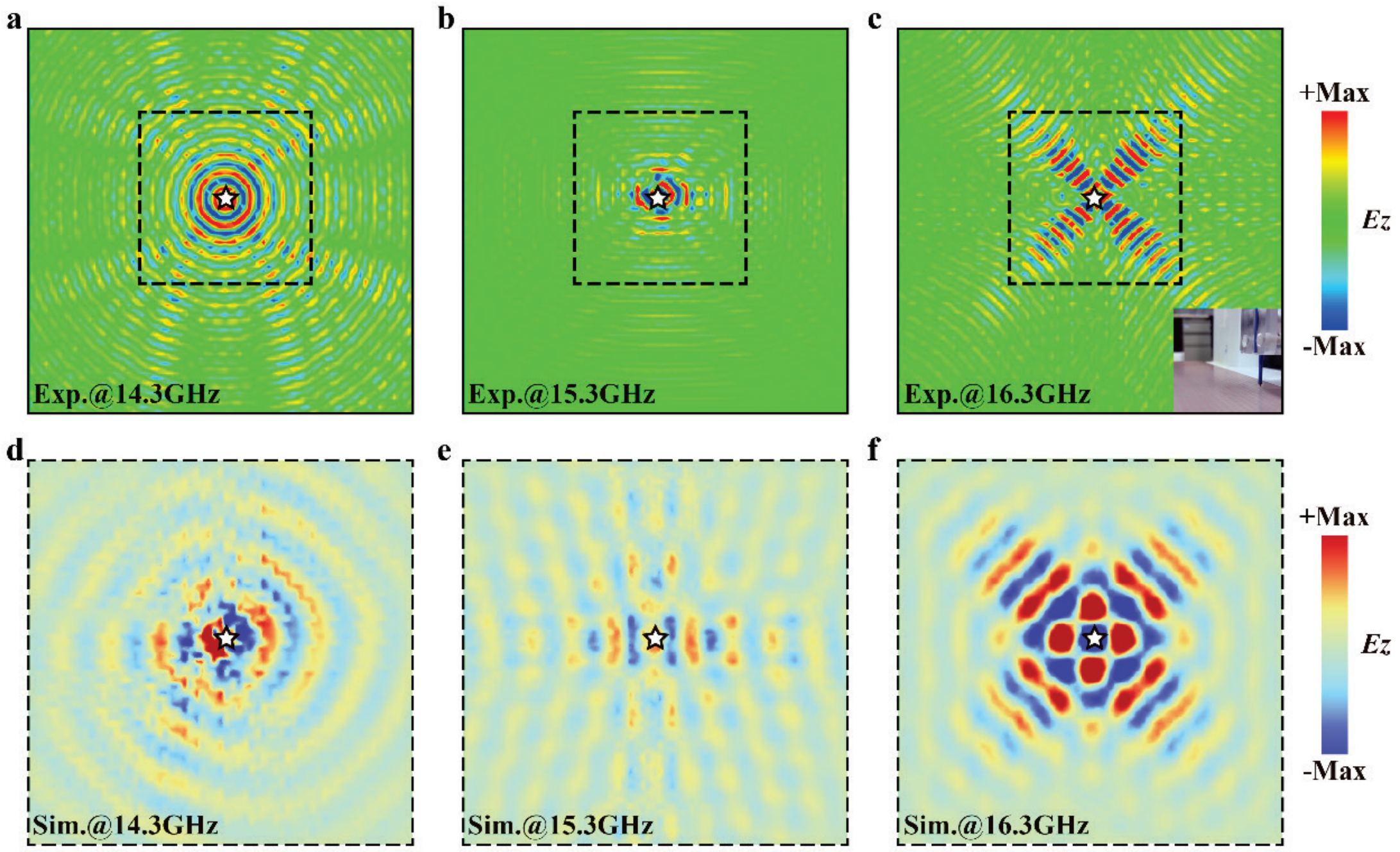

Figure 2 | Experimental verification of flatband-induced electromagnetic localization. (a–c) Measured near-field electric-field distributions of the metasurface at 14.3, 15.3, and 16.3 GHz obtained by scanning near-field microwave microscopy. The white star indicates the excitation position.(d–f) Corresponding simulated near-field distributions under identical excitation conditions. (a,d) At 14.3 GHz below the flatband, dispersive modes support extended radial wave propagation with circular interference fringes. (b,e) At 15.3 GHz, the excited flatband states exhibit suppressed lateral energy transport and strong EM localization due to near-zero group velocity. (c,f) At 16.3 GHz above the flatband, the excited dispersive modes recover directional wave propagation, resulting in anisotropic cross-shaped field distributions. The dashed red box indicates the boundary of the metasurface enlarged cell array.

### 2.3. Angular and polarization robustness of flatband-enabled Fano resonance

Following the near-field verification of flatband-induced electromagnetic localization, we investigate the far-field transmission characteristics of the $\beta$ = 90° metasurface using full-wave frequency-domain simulations. The effects of incident angle and polarization on the Fano resonance are examined to evaluate the robustness enabled by the engineered flatband and orthogonal resonator configuration. All geometric and material parameters are identical to those used in the preceding eigenmode and near-field analyses. Since the Fano resonance arises from interference between a localized resonant state and a broadband radiation channel, a stable coupling condition is essential for robust resonant responses.

We first examine polarization dependence at normal incidence. As shown in Figure 3(b), the s- and p-polarized spectra exhibit nearly identical Fano line shapes and resonance frequencies, demonstrating polarization-independent resonant responses. This robustness originates from the orthogonal arrangement of the resonators in the enlarged 2 × 2 unit cell, which supports complementary flatband eigenmodes and provides equivalent excitation pathways for the two orthogonal polarizations. Thus, the interference between the localized flatband resonance and the broadband FP radiation channel is maintained for both s- and p-polarized excitation.

We next investigate the angular dependence. Figures 3(c–e) show the simulated spectra under s- and p-polarized illumination at incident angles of 10°, 20°, and 30°, respectively. With increasing angle, the resonance frequency remains nearly unchanged, with only minor variations in amplitude and linewidth. The angular evolution is summarized in Figures 3(f,g) for s- and p-polarized excitation, respectively. Over the range of 0°–30°, the Fano resonance remains close to the flatband frequency with weak spectral variation. This angular robustness results from the weak momentum dispersion of the flatband state, which suppresses resonance shifts induced by changes in the in-plane wavevector. Meanwhile, the orthogonal resonator configuration maintains comparable coupling conditions for the two polarizations. These results demonstrate robust Fano responses against both polarization and incident-angle variations.

To further verify the physical origin and quantitative description of this robust response, we employ TCMT to model the transmission spectra. Figures 3(h,i) present the theoretical results for s- and p-polarized excitation, respectively. The TCMT calculations reproduce the main spectral characteristics of the full-wave simulations, including the nearly invariant resonance frequency and gradual evolution of the asymmetric line shape. Small discrepancies can be observed between the theoretical predictions in Figs. 3(h) and 3(i). In our analytical framework, each meta-atom is treated as an ideal lumped resonant oscillator with infinite periodic boundary approximation, neglecting two practical effects: (1) the higher-order evanescent Bloch modes excited at large incident angles inside the supercell; (2) the weak polarization cross-coupling induced by the finite geometrical asymmetry of the H-shaped meta-structure under p-polarized oblique illumination. The good agreement confirms that the robust Fano response can be consistently described as the interference between the localized flatband resonance and the broadband FP background channel.

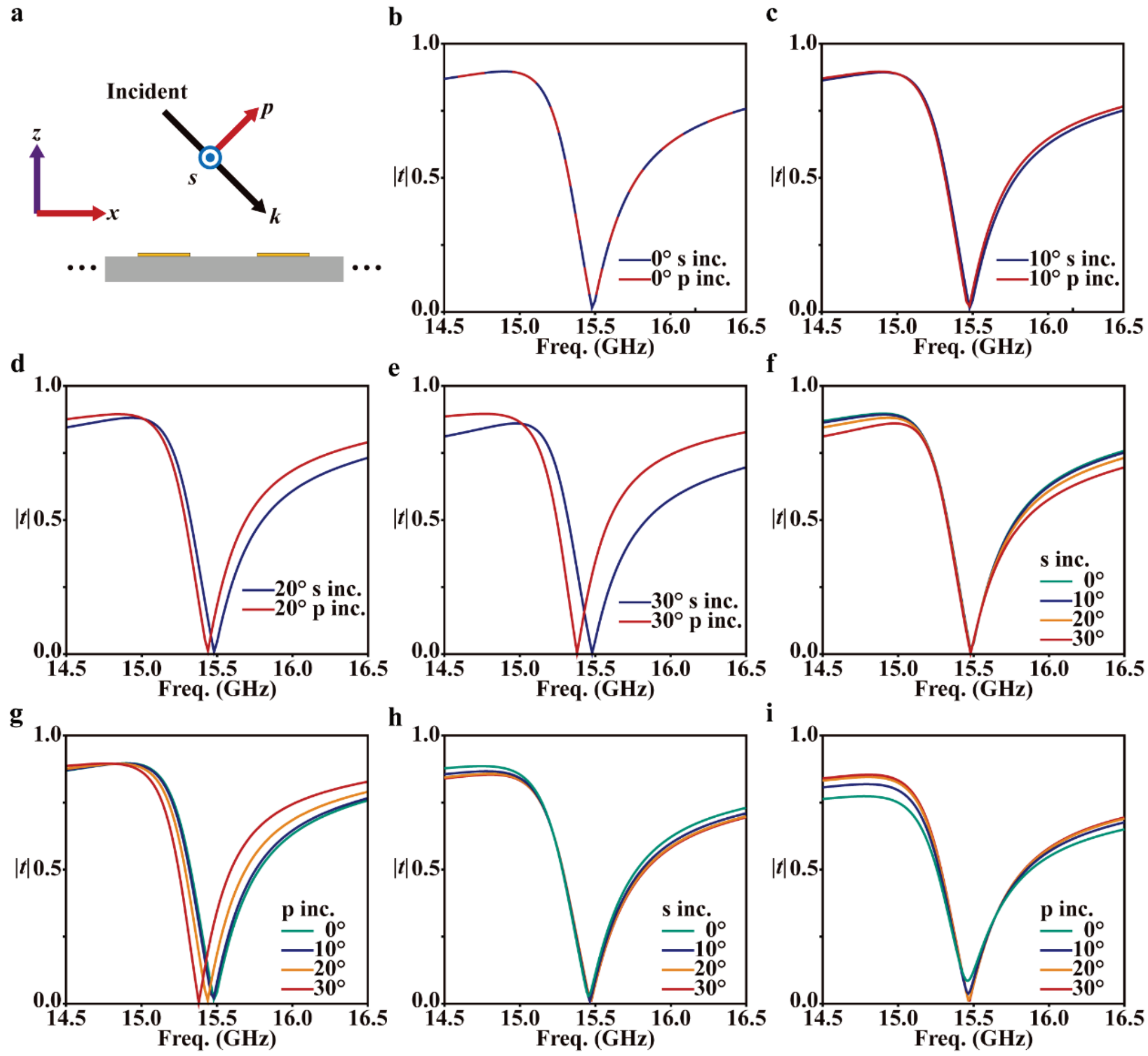

Figure 3 | SSimulated and theoretical angular and polarization robustness of the flatband-enabled Fano resonance. (a) Schematic illustration of oblique illumination with s- and p-polarized incident waves. (b) Simulated transmission spectra under s- and p-polarized illumination at normal incidence (0°). (c–e) Simulated transmission spectra under s- and p-polarized illumination at incident angles of 10°, 20°, and 30°, respectively. (f,g) Angular evolution of the simulated transmission spectra for s- and p-polarized excitation, respectively, from 0° to 30°. The Fano resonance remains nearly unchanged in frequency with only minor variations in its spectral amplitude and linewidth, demonstrating robust responses against both polarization and incident-angle variations. (h,i) Corresponding TCMT theoretical transmission spectra for s- and p-polarized excitation, respectively, reproducing the angular evolution of the simulated Fano responses.

To quantitatively describe the coherent interference between the broadband FP background channel and the discrete localized flatband resonance, we employ TCMT to establish an analytical relation between the resonant mode and the transmission response. Because the metallic resonators occupy only approximately 3% of the unit-cell area, the direct transmission of the metasurface can be approximated by that of an equivalent dielectric slab, which provides the non-resonant FP background channel. Within this framework, the transmission coefficient is written as

$$t(f, \mathbf{k}_{inc}) = t_{\mathrm{FP}}(f, \mathbf{k}_{inc}) + \frac{A\gamma_a e^{i\theta}}{f - f_0 - \mathrm{i}\gamma_a}, \quad (1)$$

where $f_0$ denotes the resonance frequency of the localized flatband mode, while $\gamma_a$ represents its amplitude decay rate. $t_{\mathrm{FP}}(f, \mathbf{k}_{\mathrm{inc}})$ describes the non-resonant Fabry–Pérot background transmission of the dielectric substrate, with $\mathbf{k}_{\mathrm{inc}} = (k_x, k_y)$ denoting the in-plane wavevector of the incident wave. Here, $A$ is the amplitude coefficient of the resonant contribution, and $e^{i\theta}$ accounts for the relative phase between the resonant scattering channel and the direct FP background channel.

The above analytical expression Eq. (1) is used to fit the theoretical transmission spectra under different incident angles and polarization states. The fitting parameters are determined independently for the corresponding simulated spectra. As shown in Figures 3(h,i), the TCMT results generally reproduce the simulated transmission spectra for s- and p-polarized excitations over the investigated angular range, capturing both the nearly invariant resonance frequency and the characteristic asymmetric Fano line shape. The good agreement between the TCMT description and full-wave simulations confirms that the observed transmission response can be consistently understood as the interference between the localized flatband resonance and the broadband FP background channel. Detailed derivation and fitting procedures are provided in Supplementary Information (Note 5).

### 2.4. Microwave Far-Field Experimental Verification of the Flatband Pattern ($\beta$=90°)

To experimentally validate the robustness predicted by the numerical simulations, far-field microwave transmission measurements are performed on the fabricated $\beta$ = 90° metasurface. A vector network analyzer (VNA)-based measurement system is employed, with the experimental configuration designed to reproduce the simulated illumination conditions and the experimental setup is shown in SI, Note 2. The polarization state is controlled by rotating the transmitting horn antenna, while the incident angle is accurately adjusted using a rotary stage over the range of 0°–30°. Before each measurement, background calibration using a bare substrate is conducted to remove system and substrate contributions. Multiple measurements are repeated and averaged to ensure experimental reliability.

Figure 4 (a–d) compare the measured transmission spectra for s- and p-polarized excitations under different incident angles. The experimentally observed Fano resonance remains centered near 15.5 GHz for all investigated polarization states and incidence conditions, in good agreement with the simulated response. Although a slight frequency offset and linewidth broadening are observed compared with simulations, these differences mainly originate from fabrication imperfections, sample alignment deviations, and unavoidable experimental losses. Importantly, the resonance evolution remains highly consistent between s- and p-polarized excitations, demonstrating the polarization robustness of the proposed metasurface. The negligible spectral difference between orthogonal polarization states confirms that the orthogonally arranged H-resonator enlarged cell provides stable excitation channels for the localized flatband modes.

The angle-dependent transmission responses are further summarized in Figure 4(e,f) for s- and p-polarized waves, respectively. Across the entire oblique incidence range from 0° to 30°, the Fano resonance dip remains tightly confined around the designed operating frequency with only minor spectral variation. The measured maximum resonance shift is approximately 0.03 GHz, while the resonance linewidth and spectral contrast remain nearly unchanged. A small reduction in resonance amplitude at larger incident angles is observed, which can be attributed to the modified coupling efficiency of the Fabry–Pérot background mode under oblique illumination. Nevertheless, the asymmetric Fano profile and resonance position are well preserved, confirming the strong angular robustness enabled by the flatband state.

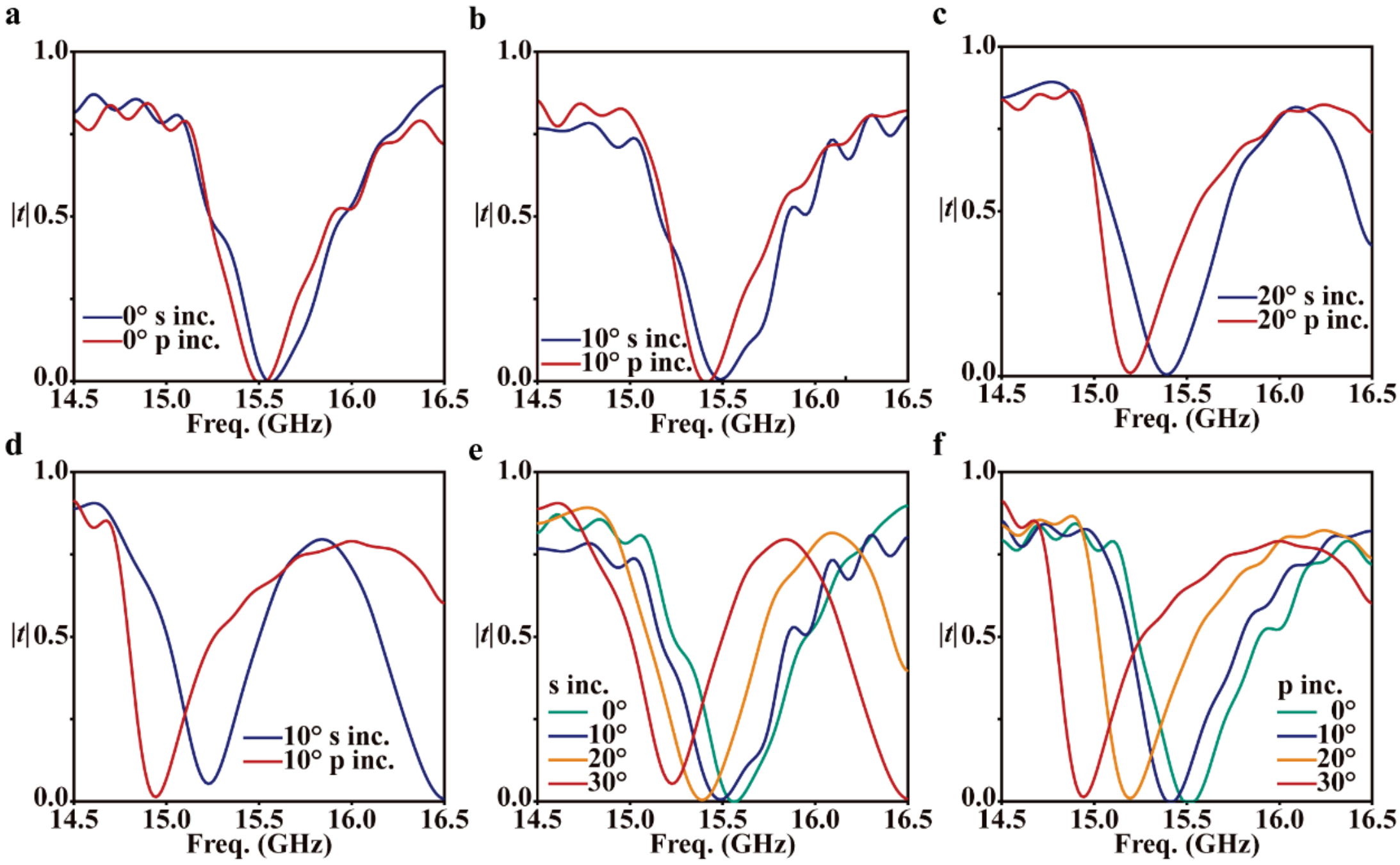

Figure 4 | Experimental verification of angular and polarization robustness of flatband-enabled Fano resonance. (a–d) Measured transmission spectra for s-polarized and p-polarized excitations at incident angles of 0°, 10°, 20°, and 30°, respectively. The Fano resonance remains stable with negligible polarization-dependent spectral variation. (e,f) Angle-dependent transmission spectra under s-polarized and p-polarized illumination, respectively. The resonance dip remains near 15.4 GHz over the 0°–30° incidence range, demonstrating experimentally verified angular and polarization robustness.

The experimental spectra exhibit small high-frequency fluctuations across the measurement range, which are mainly associated with multipath scattering and weak environmental electromagnetic interference in microwave measurements. These fluctuations do not influence the intrinsic Fano resonance characteristics, including resonance position and linewidth. The excellent agreement between simulation and experiment demonstrates that the proposed flatband engineering strategy effectively maintains stable Fano interference under practical illumination conditions.

Overall, the far-field measurements experimentally confirm the two key advantages of the proposed metasurface. The orthogonal H-resonator configuration enables robust polarization response, while the flatband-induced weak momentum dependence suppresses angular resonance variation. These results validate the physical mechanism of flatband-regulated Fano resonance and demonstrate its potential for stable microwave filtering, sensing, and resonant photonic applications.

## 3. Conclusion

In summary, we have demonstrated a Brillouin-zone-folding approach to engineer orthogonal flatbands for realizing robust Fano resonances in a planar microwave metasurface consisting of a 2×2 enlarged cell with orthogonally arranged H-shaped resonators. By introducing the enlarged cell and tuning the inter-resonator coupling through rotation, the original photonic bands are folded and reconstructed within the reduced Brillouin zone, resulting in the formation of nearly dispersionless flatbands around 15.3 GHz. The flatband states exhibit enhanced electromagnetic localization and weak wavevector dependence, providing a stable localized resonant channel for interacting with the FP radiation continuum and maintaining the asymmetric Fano interference under varying excitation conditions.

The orthogonal H-resonator configuration further enables two polarization-decoupled dipole eigenmodes, which provide comparable excitation pathways for s- and p-polarized waves and effectively suppress polarization-induced spectral variations. Near-field microwave measurements directly verify the transition from dispersive wave propagation to flatband-induced electromagnetic localization, confirming the physical origin of the engineered flatband states. Furthermore, far-field transmission simulations and experiments demonstrate that the Fano resonance remains stable under oblique illumination from 0° to 30° for both polarization states, with only minor resonance shifts and amplitude variations.

The deviations between simulations and experiments, including resonance-frequency offsets and linewidth broadening, mainly originate from fabrication imperfections, sample alignment errors, and experimental losses, while the essential flatband-mediated Fano coupling mechanism remains unaffected. Compared with conventional FP–Lorentz coupled Fano metasurfaces, whose resonant responses are strongly influenced by incident wavevector and polarization conditions, the proposed design achieves enhanced resonant robustness by simultaneously controlling photonic dispersion and resonator symmetry. These results establish a direct connection between band-folding-induced flatbands and stable Fano interference, providing a practical route for designing robust planar metasurface resonators.

Looking forward, this band-folding flatband strategy can be readily extended to higher operating frequencies spanning terahertz and infrared regimes by downscaling the meta-atom dimensions, where angle- and polarization-insensitive resonant devices are in high demand for sensing, filtering, and optical signal processing applications. Further opportunities also lie in integrating active tuning mechanisms, such as phase-change materials or electrically tunable components, into the enlarged supercell architecture to realize reconfigurable flatband-Fano responses. Additionally, combining flatband engineering with topological metasurface concepts may unlock new classes of resilient resonant modes that are tolerant against structural defects, paving the way toward high-performance, robust photonic metadevices under complex real-world illumination environments.

**Acknowledgement**

This research was supported by the National Natural Science Foundation of China (No. 12304348), Guangdong Basic and Applied Basic Research Foundation (No. 2025A1515011470), Guangdong Provincial Project (No. 2023QN10X059), Guangzhou-HKUST(GZ) Joint Funding Program (2025A03J3783).

**Data Availability Statement**

The data that support the findings of this work are available from the corresponding authors upon reasonable request.

**Supplemental Materials**

See the Supplemental Material for detailed information regarding the COMSOL simulation configurations, full evolution of photonic band structures and density-of-states spectra under varied resonator rotation angle $\beta$, dual microwave experimental setups for near-field scanning and far-field transmission measurements, as well as the full derivation of the temporal coupled-mode theory combined with the transfer-matrix method that describes the Fano interference between flatband resonant modes and Fabry-Pérot background continua.